\documentclass[letterpaper]{jpconf}
\usepackage{graphicx,epsfig,latexsym,color}
\begin{document}

\title{Analysis of a Material Phase Shifting Element in an Atom Interferometer}
\author{John D.\ Perreault and Alexander D.\ Cronin}
\address{University of Arizona, Tucson, Arizona 85721, USA}
\ead{johnp@physics.arizona.edu}

\begin{abstract}
The interaction of Na atoms with a surface was probed by inserting
a nanofabricated material grating into one arm of an atom
interferometer (IFM).  This technique permits a direct measurement
of the change in phase and coherence of matter waves as they pass
within 25 nm of the grating bar surface.  The practical concerns
and challenges of making such a measurement are discussed here.
Interference of spurious diffraction orders, IFM path overlap, and
the partial obscuration of IFM beams are all important aspects of
this experiment.  The systematic effects that contribute to the
measured phase shift and contrast are discussed.
\end{abstract}

\begin{figure}[t]
\scalebox{.75}{\includegraphics{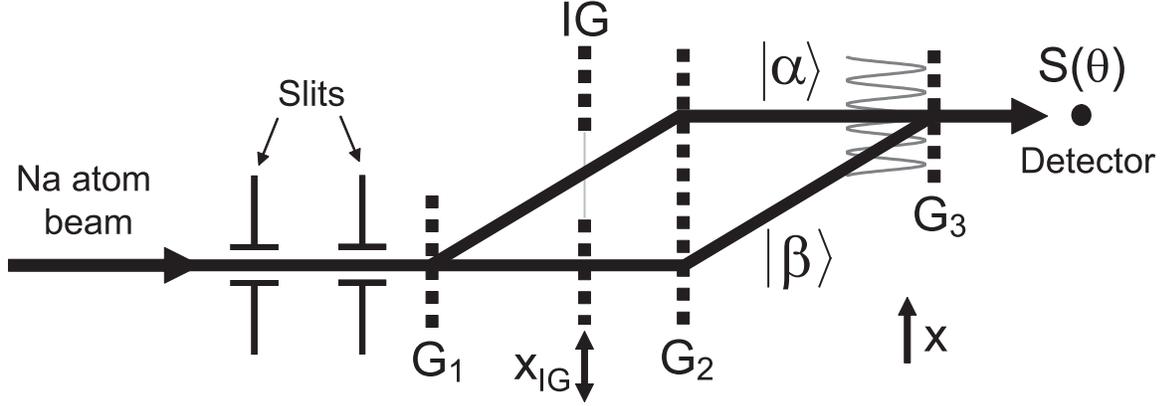}}
\caption{\label{fig:setup}Experimental setup for using an atom
interferometer (IFM) to measure the vdW induced phase shift.  An
IFM is formed using the zeroth and first order diffracted beams of
gratings $G_{1}$ and $G_{2}$.  Placing the interaction grating
(IG) in one of the interferometer paths causes the interference
pattern at the third grating $G_{3}$ to shift in space. The IG has
been perforated (light grey line) to allow the reference arm of
the IFM to pass unaffected. The flux transmitted through $G_{3}$
is the detector signal $S(\theta)$.}
\end{figure}

Atomic diffraction from material grating structures \cite{sava96}
has been used as a tool to measure atom-surface interactions for
noble gases \cite{gris99} and alkali atoms \cite{cron04,perr05}.
In these experiments the van der Waals (vdW) interaction
\cite{milo94} changed the relative intensities of the diffraction
orders.  More recently the atom wave phase shift $\Phi_{0}$
induced by these grating structures\footnote{The subscript of
$\Phi_{0}$ specifies this variable as the phase of the zeroth
diffraction order induced by the IG.  For reasons discussed in
\cite{pifm05} only the zeroth diffraction order leads to
significant interference contrast.} was measured directly using a
sodium atom beam interferometer (IFM) \cite{pifm05}. The
experimental setup is shown in Fig. \ref{fig:setup}.  An
interaction grating (IG), comprised of an array of 50 nm wide
channels, is inserted into one arm of the IFM. The vdW interaction
between the sodium atoms and IG causes a phase shift of about
$\Phi_{0}\sim 0.3$ radians. A number of systematic effects need to
be understood before reporting this phase shift induced by the vdW
interaction. This paper describes how the measured phase shift and
contrast in \cite{pifm05} are influenced by these systematic
effects and suggests some physical mechanisms for them.

\begin{figure}[t]
\begin{minipage}{18pc}
\includegraphics[width=18pc]{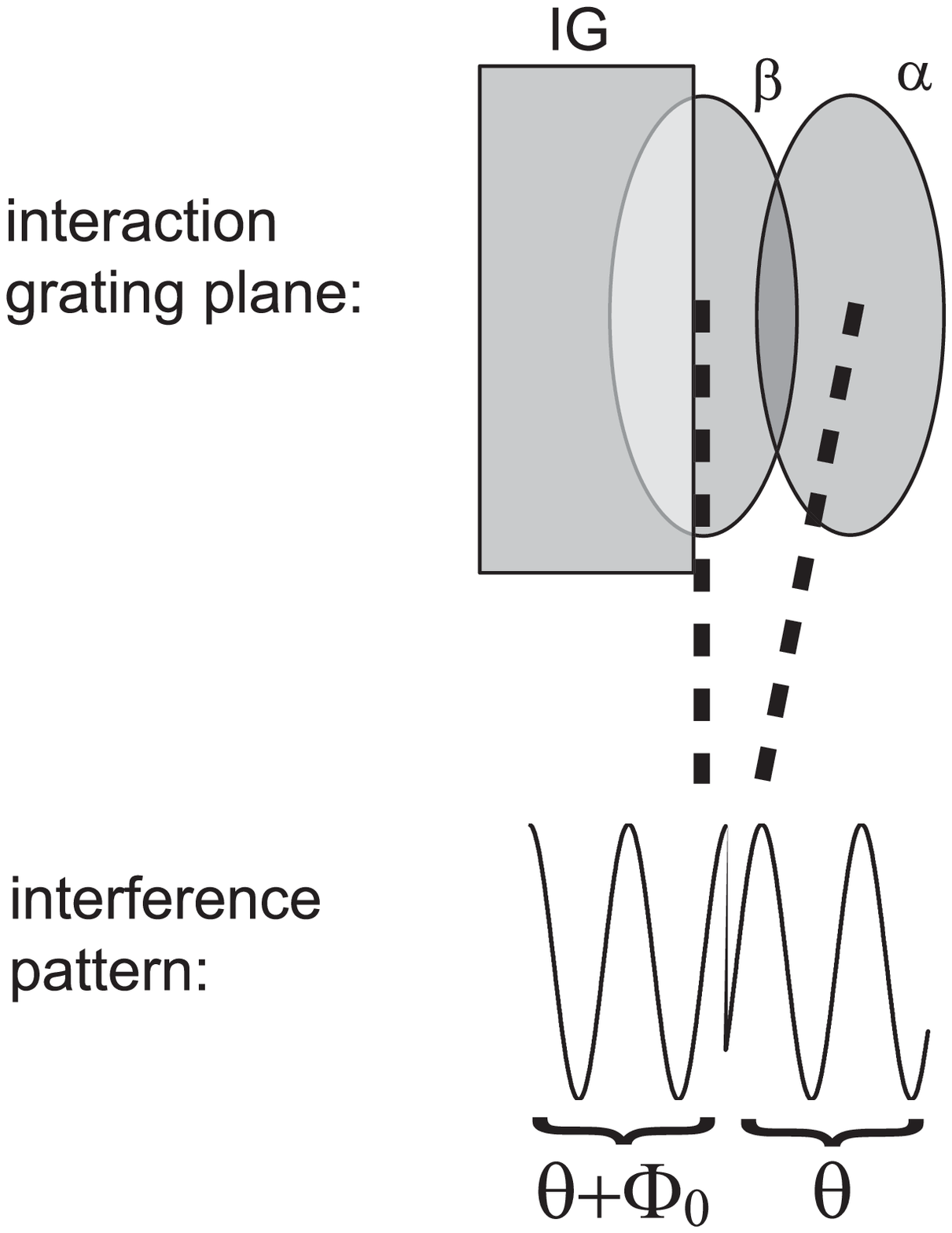}
\caption{\label{fig:obscure}Partial obscuration of the
interferometer beams $\alpha$ and $\beta$.  In general the
interaction grating (IG) may only induce a phase shift to part of
the beam, resulting in an interference pattern that has different
phases in different regions of space.  As indicated by the dark
grey region the beams can also have some overlap resulting in a
more complicated interference pattern.}
\end{minipage}\hspace{2pc}%
\begin{minipage}{18pc}
\includegraphics[width=18pc]{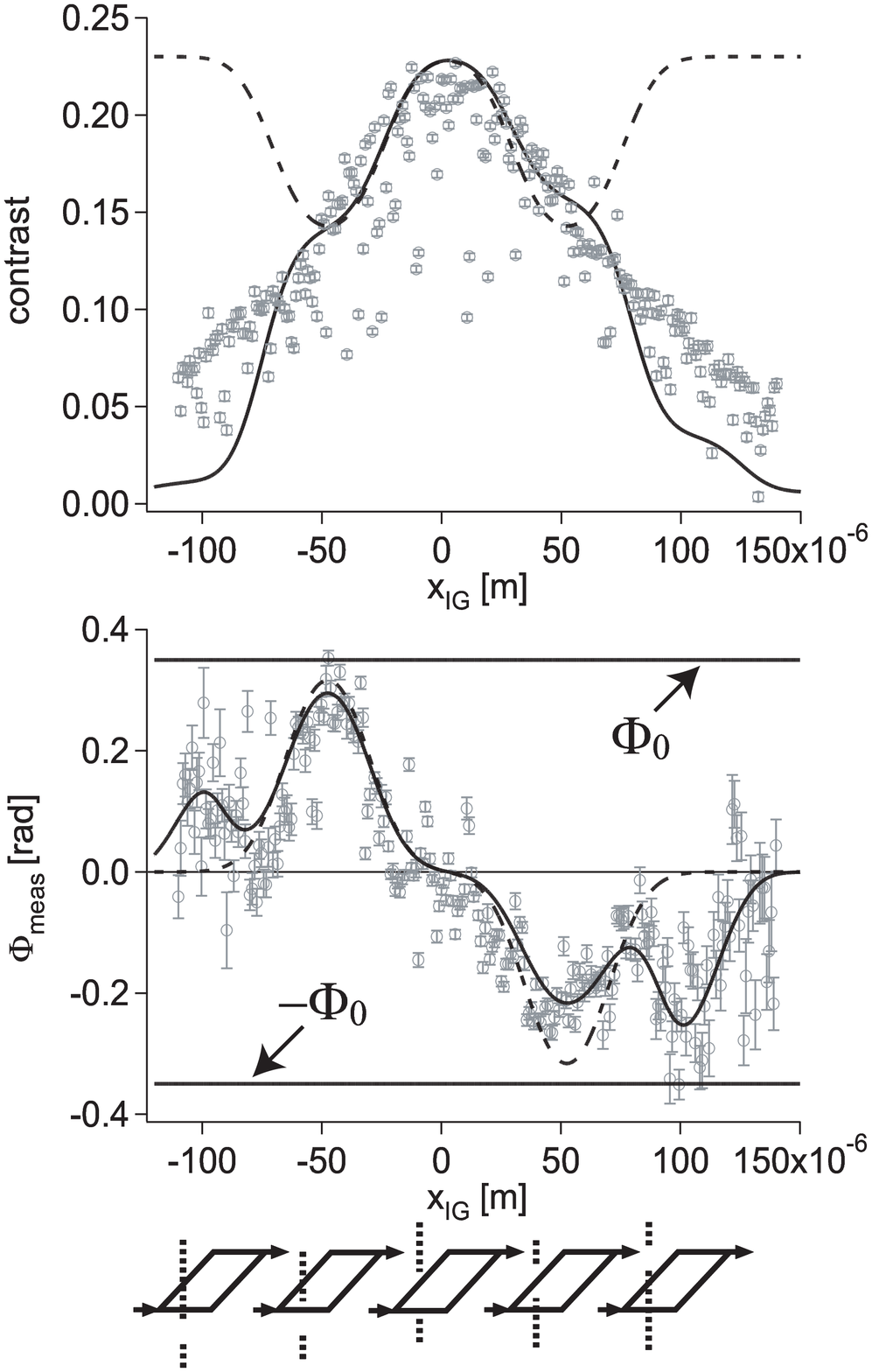}
\caption{\label{fig:profdata}Measured phase and contrast as a
function of interaction grating position $x_{IG}$.  The naive
model implied by Fig. \ref{fig:setup} (dashed) fails to reproduce
several features in the phase and contrast, which are explained
when systematic effects are included (solid).  The horizonal lines
on the phase plot indicate the value for the induced phase
$\Phi_{0}$ and the diagrams below the x-axis show the position of
the IG within the interferometer.}
\end{minipage}
\end{figure}

In general there is a non-trivial relationship between the
measured phase shift $\Phi_{meas}$ and induced phase shift
$\Phi_{0}$, when the IFM paths are partially obscured by a phase
shifting element. Therefore, care should be taken when
interpreting the phase shift data.  This notion of partial
obscuration is shown in Fig. \ref{fig:obscure}, which illustrates
how the interference pattern can have different phases in
different regions of space. The detected interference signal can
be written as the average flux transmitted through each grating
window of $G_{3}$ in Fig. \ref{fig:setup}
\begin{eqnarray}
S(\theta) & = & \sum_{l}\frac{1}{d}\int_{-w/2}^{w/2}dx\left[1 +
C(x-ld)\cos\left(\frac{2\pi}{d}(x-ld) + \theta +
\phi(x-ld)\right)\right]\langle I\rangle(x-ld)\nonumber\\ &
\approx & \frac{1}{d}\sum_{l}\int_{-w/2}^{w/2}dx\left[1 +
C_{l}\cos\left(\frac{2\pi}{d}x + \theta +
\phi_{l}\right)\right]\langle I_{l}\rangle \nonumber\\ & = &
\frac{w}{d}\sum_{l}\left[1 +
\mbox{sinc}\left(\frac{w}{d}\right)C_{l}\cos(\theta +
\phi_{l})\right]\langle I_{l}\rangle,
%S(\theta) =  \frac{w}{d}\sum_{n}\left[1 + C_{n}\cos(\theta +
%\phi_{n})\right]\langle I_{n}\rangle,
\label{eq:signal}
\end{eqnarray}
where the contrast ($C_{l}$), phase ($\phi_{l}$), and average
intensity ($\langle I_{l}\rangle$) of the interference pattern are
assumed to be constant over each grating window $l$ of $G_{3}$.
The grating window size $w$ and period $d$ in Eqn. \ref{eq:signal}
refer to grating $G_{3}$.\footnote{A more complete discussion of
the grating structures and other experiments in which they are
used can be found in a separate entry of the CAMS conference
proceedings by A. D. Cronin.} The variable $\theta=2\pi x_{3}/d$
accounts for the position of $G_{3}$ relative to the interference
pattern phase $\phi_{l}$. Equation \ref{eq:signal} establishes a
connection between the spatial interference pattern, shown just
before $G_{3}$ in Fig. \ref{fig:setup}, and the signal $S(\theta)$
which is actually measured.

From Eqn. \ref{eq:signal} it is clear that the detector signal
$S(\theta)$ is a sum of cosines with varying phases $\phi_{l}$ and
intensities $C_{l}\langle I_{l}\rangle$.  When determining the
measured phase $\Phi_{meas}$ of the signal it is only the relative
phase and intensity of the terms in Eqn. \ref{eq:signal} that are
important. For the case of a half-plane phase shifting element (e.
g. the IG in Fig. \ref{fig:obscure}) the form for the detector
signal implied by Eqn. \ref{eq:signal} would be
\begin{equation}
S(\theta) \propto A\cos(\theta) + B\cos(\theta + \Phi_{0}) \equiv
D\cos(\theta + \Phi_{meas}), \label{eq:signalhalf}
\end{equation} where constant offsets in the signal have been
ignored.  The intensity of the detector signal can then be given
by
\begin{equation}
D = \sqrt{A^{2} + B^{2} + 2AB\cos(\Phi_{0})},
\label{eq:mintensity}
\end{equation}
and the phase by
\begin{equation}
\Phi_{meas} = \tan^{-1}\left[\frac{B\sin(\Phi_{0})}{A +
B\cos(\Phi_{0})}\right], \label{eq:mphase}
\end{equation}
where $A$ and $B$ are the relative intensities of the unshifted
and shifted interference patterns.  The resulting phase and
contrast measured by $S(\theta)$ can also be found for more
complicated interference patterns by using Eqn.
\ref{eq:signalhalf} in an iterative fashion.

Figure \ref{fig:profdata} shows the measured phase and contrast of
$S(\theta)$ when the interaction grating is placed at a given
location $x_{IG}$ inside the atom IFM.  All phase measurements are
relative to the situation where the IG is out of the IFM, and this
reference phase was regularly measured. As one would expect the
measured phase reaches a local extremum when the IG is completely
obscuring one of the IFM paths.  When the IG begins to obscure
both of the IFM paths one would presume that the measured phase
should return to zero again.  However, the $\Phi_{meas}$ data
deviate from this prediction.  Likewise, the contrast should
decrease when the IG attenuates one of the IFM paths, but then
return to its nominal value when both paths are obscured.  In
addition, the beam overlap shown in Fig. \ref{fig:obscure} will
tend to make the measured phase smaller than the induced phase
because the overlapped portion will have no relative phase
difference.  These expectations are made quantitative by the use
of Eqn. \ref{eq:signalhalf} and shown as the dashed line in
\mbox{Fig. \ref{fig:profdata}}.  There are two striking failures
of this naive prediction when compared to the experimental data:
the appearance of a phase shift and significant loss of contrast
when \emph{both} IFM paths are obscured by the IG.

\begin{figure}[t]
\begin{minipage}{18pc}
\includegraphics[width=18pc]{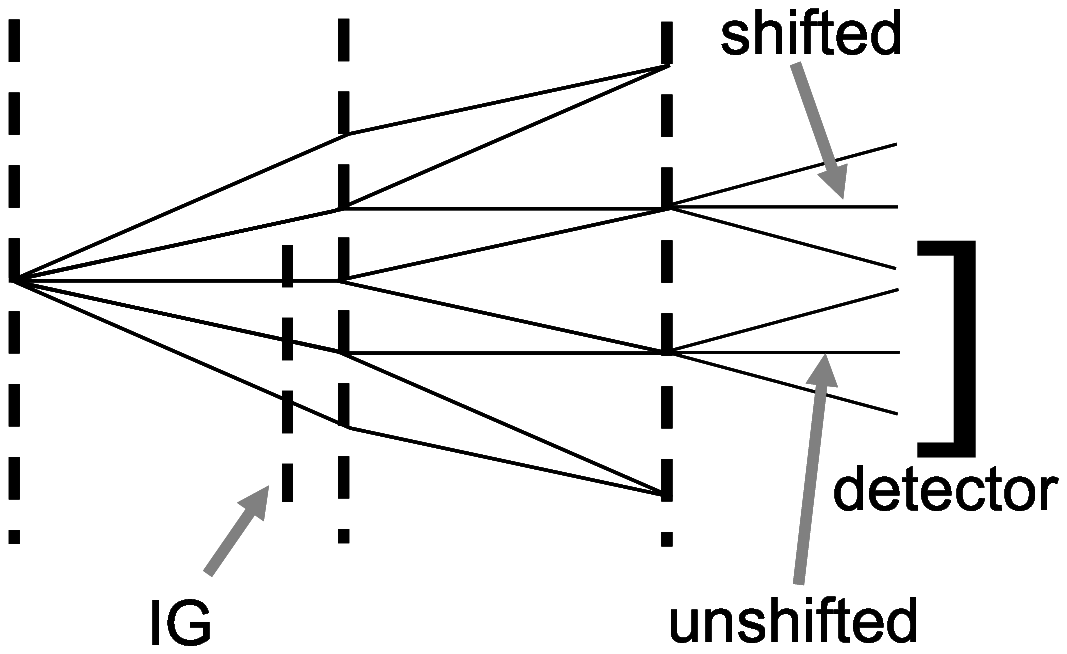}
\caption{\label{fig:multifm}Explanation for the observed phase
shift when both interferometer paths are obscured by the IG.  The
other diffraction orders can interfere leading to additional
contributions to $S(\theta)$.  It is possible for the IG to
obscure both of the primary interferometer paths while only
obscuring one of the secondary paths, leading to a phase shifted
component.}
\end{minipage}\hspace{2pc}%
\begin{minipage}{18pc}
\includegraphics[width=18pc]{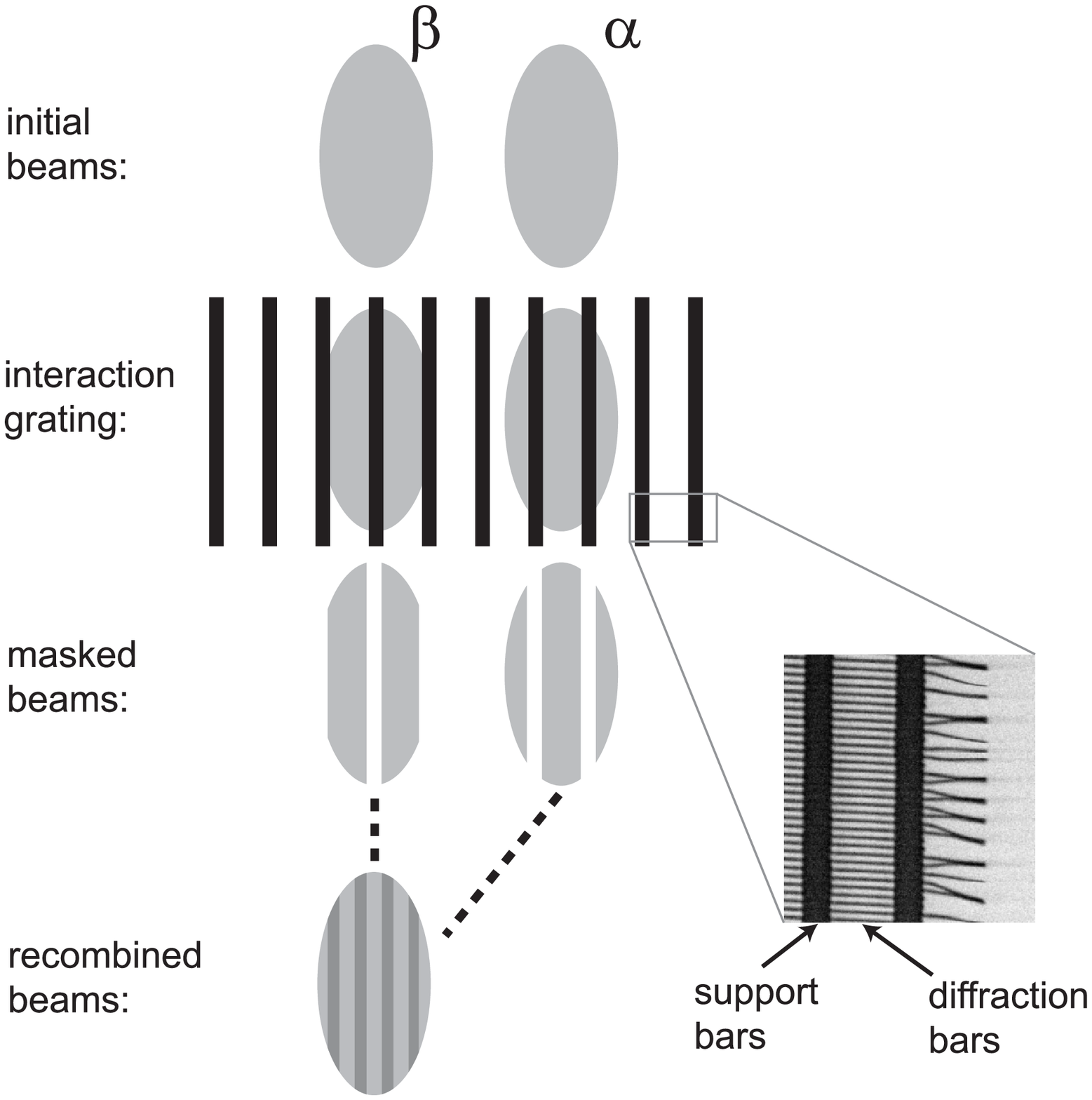}
\caption{\label{fig:masking}Explanation for contrast reduction
when the IG obscures both interferometer paths. The IG support
bars effectively operates like a mask.  When the beams are
recombined only the dark grey area will have significant contrast,
leading to an overall reduction in the observed contrast in
$S(\theta)$.  The inset SEM image shows how the IG has been
prepared with a sharp transition from gap to intact grating.}
\end{minipage}
\end{figure}

The appearance of the extra phase features can be understood by
looking more carefully at the details of the IFM.  Our IFM is
formed by the zeroth and first order diffracted beams of the
gratings $G_{1}$ and $G_{2}$ in Fig. \ref{fig:setup}. In reality
there are more than just two paths that can interfere because of
the other diffraction orders. This situation is depicted in Fig.
\ref{fig:multifm}. These additional interfering paths allow for
the possibility of the IG to obscure both of the primary
interferometer paths, while only obscuring one of the secondary
interferometer paths. When this notion is combined with the finite
size of the detector and diffraction caused by the third grating
$G_{3}$, a clear mechanism for the extra phase features is found.

A likely explanation for the unexpected reduction in contrast,
when both interferometer paths are obscured by the IG, is shown in
Fig. \ref{fig:masking}.  The inset SEM image in Fig.
\ref{fig:masking} shows how the diffraction bars (which cause the
atom-surface phase shift) are stabilized by much more widely
spaced support bars.  The support bars will imprint a spatial
amplitude modulation on the two beams, shifted in space by
different amounts with respect to the center of the beams. When
the beams are recombined the region of overlap is effectively
reduced, leading to an overall reduction in contrast. While there
are some near-field diffractive effects caused by the support bars
\cite{good96}, numerical simulations have shown that the effective
mask picture in Fig. \ref{fig:masking} is still appropriate when
considering the influence on $S(\theta)$.  It is important to note
that this effect only reduces the contrast for an IFM that has
\emph{both} paths obscured by the IG. This also explains the
relative prominence of the extra phase features, since the
contrast of the primary interferometer is reduced compared to the
secondary one as a result of this effect.

When the previously discussed systematic effects are incorporated
into the coefficients $A$ and $B$ in Eqn. \ref{eq:signalhalf},
much better agreement with the data is achieved.  The solid line
in Fig. \ref{fig:profdata} shows the prediction of a model which
includes the influence of other interfering orders and the support
bars.  It is quite satisfying to see that the behavior of the
measured contrast and phase is now understood even when the IG is
blocking both of the primary interferometer paths.  One can also
see that the asymmetry of the phase profile is reproduced.

In conclusion the measured phase and contrast as a function of IG
position are now understood to be influenced by a number of
systematic effects. The primary physical mechanisms for the
systematic effects are beam overlap, interference of additional
diffraction orders, and an effective masking by the IG support
bars.  The inclusion of systematic effects leads to a relationship
between the phase shift that is measured ($\Phi_{meas}$) with our
experiment and that which is actually induced ($\Phi_{0}$) by the
IG. This allows us to make quantitative comparisons to predictions
for the phase shift $\Phi_{0}$ and in turn the vdW coefficient
$C_{3}$ \cite{pifm05}.

\ack This research was supported by grants from Research
Corporation and the National Science Foundation.

\section*{References}
%\bibliography{vprof}

\end{document}